\documentstyle[12pt,amssymb]{article}
\input epsf
\def\theequation{\arabic{section}.\arabic{equation}}
\newcounter{subequation}[equation]
\makeatletter
\@addtoreset{equation}{section}
\expandafter\let\expandafter\reset@font\csname reset@font\endcsname
  {\endeqnarray\stepcounter{equation}}
\makeatother
\newcommand{\be}{\begin{equation}}
\newcommand{\ee}{\end{equation}}

\normalbaselines
\begin{document}
\title{Possible confinement mechanisms for nonrelativistic  particles on a line}
\author{
P.C. Stichel\\
An der Krebskuhle 21\\ D-33619 Bielefeld, Germany \\
e-mail:pstichel@gmx.de
\\ \\
W.J. Zakrzewski
\\
Department of Mathematical Sciences, University of Durham, \\
Durham DH1 3LE, UK \\
 e-mail: W.J.Zakrzewski@durham.ac.uk\\
and\\
Center for Theoretical Physics\\
Massachusetts Institute of Technology, 
Cambridge, 02139, USA}
\date{}
\maketitle

\begin{abstract}
The gauge model of nonrelativistic particles 
on a line interacting with nonstandard
gravitational fields [5] is supplemented by the addition of  a
(non)-Abelian gauge interaction. Solving for the gauge fields
we obtain equations, in closed form, for a classical
two particle system. The corresponding 
 Schr\"{o}dinger equation, obtained by the Moyal quantization
procedure, is solved analytically. Its solutions exhibit two different
confinement mechanisms - dependent on the sign of the coupling
$\lambda$ to the nonstandard gravitational fields. For $\lambda>0$
confinement is due to a rising potential whereas for $\lambda<0$ it is due to
to the dynamical (geometric) bag formation. Numerical results for the corresponding energy spectra are given. For a particular relation between
two coupling constants the model fits into the scheme of supersymmetrical
quantum mechanics.
\end{abstract}

\section{Introduction}

Two dimensional models have often been considered as a testing laboratory for various
ideas in elementary particle theory (\cite{one}).
One of the most outstanding problems in elementary particle theory is the nature
of quark confinement, which is a non-perturbative phenomenon. Many models of it have 
been presented (see (\cite{two}) and the literature cited therein) but, so far, there is 
no understanding of it in terms of basic principles. In the framework of  two-dimensional
gauge theories {\it ie}
$QED_2$ (\cite{three}) and $QCD_2$ (\cite{four}) confinement is due to a linearly
rising potential between two static fermions. This suggests that confinement should
be considered a low-energy nonrelativistic phenomenon.

In a recent paper (\cite{five}) one of the present authors (PS), considered a theory invariant
under local time-dependent nonrelativistic 1-d space translations:
\be
x\, \rightarrow \, x'(x,t).
\ee 

The corresponding particle action has been made invariant with respect to (1.1) by introducing
two gauge fields\footnote{Throughout this paper we use the notation of [5].} $h(x,t)$ and $e(x,t)$ which couple via
 a Maxwell-like interaction (\cite{five})
\be
L_{field}\,=\, {1\over 2\lambda}\,\int dx\,h(x,t)\,F\sp2(x,t),
\ee
where $F$ is an invariant field strength
\be
F:=\,{1\over h}(\partial_t h-\partial_x e)
\ee
and $\lambda$ is a coupling constant.

Note that due to the similarity of this interaction to the zweibein formalism and torsion tensors in
General Relativity (\cite{six}) we can consider this interaction as describing a nonstandard gravity.

The minimal coupling of the zweibein fields $h$ and $e$ to $N$ nonrelativistic classical particles
on trajectories $x_{\alpha}(t)$, ($\alpha=1,..N)$ is given, in the first order Lagrangian 
formalism (\cite{five}) by
\be
L_{part,0}\sp{(N)}\,=\,\sum_{\alpha=1}\sp{N}\,\xi_{\alpha}\,(h_{\alpha}\dot x_{\alpha}\,+\,e_{\alpha})\,-\,{1\over 2}\,\sum_{\alpha=1}\sp{N}\, \xi_{\alpha}\sp2.
\ee

In (\cite{five}) it was shown that the classical dynamics described by (1.2) and (1.4) leads, for
$\lambda<0$, to the phenomenon of geometric bag formation and, therefore, to confinement
in the case of two or three particles ($N=2$ or 3).

The corresponding stationary Schr\"odinger equation describing the relative motion of two such
particles is given by (\cite{five})
\be
E\,\xi_2(x)\,=\,\left[-\hbar\sp2\,\partial_x\,
\left(1+\frac{\lambda}{2}\vert x\vert\right)\partial_x
\,-\,\frac{\hbar\sp2}{4}\lambda \delta(x)\right]\,\xi_2(x)
\ee
exhibiting a singularity of the metric at $\vert x\vert=\frac{2}{\vert \lambda\vert}$ for $\lambda<0$,
{\it ie} confinement within the bag $\left[-\frac{2}{\vert\lambda\vert},\,\frac{2}{\vert\lambda\vert}\right]$.
Numerically determined values for the corresponding energy levels were given in (\cite{five}).

One of the aims of this paper is to study the extension of the theory described in (\cite{five})
by supplementing it with an additional (Non)-Abelian gauge interaction. To do this, in the gauge
sector we consider the effects of the well-known Maxwellian actions:
\begin{itemize}
\item Abelian case ($A_{\mu}$ - electromagnetic potential)
\be
S\sp{A}\,=\,{1\over 2}\,\int\,dt\,dx\,{1\over h}\,E\sp2
\ee
with the electric field $E$ given by
\be
E:=\,\partial_x\,A_0\,-\,\partial_t\,A_1
\ee
\item Non-Abelian case ($A_{\mu}\sp{i}$-isospin gauge field potential; for simplicity we consider
$SU(2)$ as the internal symmetry group)
\be
S\sp{NA}\,=\,{1\over 2\kappa}\,\int\,dt\,dx\,{1\over h}\,(E\sp{a})\sp2
\ee
with the non-Abelian electric field $E\sp{a}$ given by
\be
E\sp{a}:=\,\partial_x\,A_0\sp{a}\,-\,\partial_t\,A_1\sp{a}\,+\,\epsilon_{abc}\,A_1\sp{b}\,A_0\sp{c}.
\ee
\end{itemize}
Note that the factor $\frac{1}{h}$ in front of $(E)\sp2$ in (1.6) and (1.8), arises from the requirement
of the invariance of the action under local translations (1.1), when we have assumed that the gauge
fields transform covariantly.

The plan of this paper is as follows: In section 2 we extend the model of  (\cite{five})
(given by (1.2) and (1.4)) by adding to it the coupling to the (1+1)-dimensional electrodynamics
and discuss its classical dynamics. Corresponding results for the non-Abelian case are given in section 3.
In section 4 we describe two-body quantum mechanics, discuss the confinement mechanisms 
for both signs of the coupling constant $\lambda$ and present some numerical results for
the corresponding energy spectra. Section 5 contains some final remarks.

\section{Classical Dynamics for the Abelian case}

To describe $N$ nonrelativistic charged particles interacting with zweibein fields $e$ and $h$, and an
electromagnetic field $A_{\mu}$
we consider the following action:
\be
S_{part}\sp{(N)}\,=\,S_{part,0}\sp{(N)}\,+\,\int\,dt\,
\sum_{\alpha}\,g_{\alpha}\,(\dot x_{\alpha}\,A_{1,\alpha}\,+\,A_{0,\alpha})
\ee
where $g_{\alpha}$ is the electric charge of the $\alpha$-th particle.
Clearly, $S_{part}\sp{(N)}$ is invariant under local translations (1.1).

The full action is given by
\be
S\sp{(N)}\,=\,S_{field}\,+\,S\sp{A}\,+\,S_{part}\sp{(N)}.
\ee

Variation of $S\sp{(N)}$ with respect to the zweibein fields $h$ and $e$ gives the equations of motion (EOM)
\be
\frac{1}{\lambda}\partial_tF\,+\,\frac{1}{2\lambda}F\sp2\,+\,\frac{1}{2}{E\sp2\over h\sp2}\,=\,\sum_{\alpha}
\,\dot x_{\alpha}\,\xi_{\alpha}\,\delta(x-x_{\alpha})
\ee
and the Gauss constraint
\be
\partial_x F\,=\,-\lambda\,\sum_{\alpha}\,\xi_{\alpha}\,\delta(x-x_{\alpha}),
\ee
respectively, where $\xi_{\alpha}$ is given by the constraint (\cite{five})
\be
\xi_{\alpha}\,=\,\dot x_{\alpha}h_{\alpha}\,+\,e_{\alpha}.
\ee

In this derivation, we have assumed that $e$ and $h$ are finite at spatial infinity and that $F$ vanishes there (\cite{five}).
This has assured vanishing boundary terms (in the variations of (1.2)), the finiteness of the integral (1.2)
and it also leads to the constraint (\cite{five})
\be
\sum_{\alpha}\,\xi_{\alpha}\,=\,0.
\ee

As discussed in (\cite{five}) it is convenient to fix the gauge of the zweibein by imposing
\be
h(x,t)\,=\,1
\ee
and so, in the remainder of this paper, we work in this gauge.

As shown in (\cite{five}) the Gauss constraint (2.4) can then be solved and we have
\be
e(x,t)\,=\,\frac{\lambda}{2}\,\sum_{\alpha}\,\xi_{\alpha}(t)\,\vert x-x_{\alpha}(t)\vert\,-\,v(t).
\ee

The EOM, the Gauss constraint for the electric field E and its solutions are well known from the
(1+1) electrodynamics. Thus, in the axial gauge $A_1=0$ , we have
\be
A_0(x,t)\,=\,\frac{1}{2}\,\sum_{\alpha}\,g_{\alpha}\,\vert x-x_{\alpha}(t)\vert.
\ee

Note that the existence of (1.6) imposes the condition that $E$ vanishes at spatial infinity, and so, due to 
(2.9) giving us the constraint 
\be
\sum_{\alpha}\,g_{\alpha}\,=\,0.
\ee

This corresponds to confinement of single charged states in $QED_2$ [3]. The electric field is non vanishing only in the space between the charged
particles, {\it i.e.} it links them.

Finally, the variation of the action with respect to $x_{\alpha}$ gives the particle EOM, namely
\be
\dot \xi_{\alpha}\,+\,\xi_{\alpha}\,F_{\alpha}\,-\,g_{\alpha}\,E_{\alpha}\,=\,0.
\ee

Note that when (2.6), (2.8) and (2.11) 
are satisfied, the EOM (2.3) is automatically satisfied too.

Applying the Legendre transformation to the Lagrangian (2.2) using (2.8),
 (2.9) and the constraints (2.6) and (2.10) we find that the two-body Hamiltonian $H$, describing the relative motion, is given by
\be
H\,=\,\frac{1}{4}\,{\dot x\sp2\over 1+\frac{\lambda}{2}\vert x\vert}\,+\,\frac{g\sp2}{2}\vert x \vert ,
\ee
where we have defined
\be
x:=\,x_1\,-\,x_2
\ee
and
\be 
g:=\,g_1\,=\,-\,g_2.
\ee

We note that the electromagnetic interaction adds just the well known potential term
$\frac{g\sp2}{2}\vert x\vert$ to the Hamiltonian given 
in (\cite{five}).

From (2.12) we conclude that
\begin{itemize}
\item
For $\lambda<0$ the relative particle motion is confined to the bag $\left[-\frac{2}{\vert\lambda\vert},\,\frac{2}{\vert\lambda\vert}\right]$,
\item 
For $\lambda>0$ we have bounded motion due to the rising potential term.
\end{itemize}

\section{Classical Dynamics in the Non-Abelian case}

In the non-Abelian case the corresponding charge space coordinates are given by the
isovectors $\{Q_{\alpha}\sp{i}\}_{i=1}\sp{3}$ which, after quantization, satisfy
the equal-time commutation relations 
\be
[\hat Q\sp{i},\,\hat Q\sp{j}]\,=\,i\,\hbar\,\epsilon\sp{ijk}\,\hat Q\sp{k}.
\ee
Note that in the classical case the $\{ Q\sp{i}\}$ can be considered as vectors
on the sphere $S\sp2$ of radius $J$. So, taking on $S\sp2$ spherical coordinates
$\theta$ and $\varphi$ we get, for the corresponding part of the particle action
\be
S_{SU(2)}\sp{(N)}\,=\,\int \, dt\,\sum_{\alpha=1}\sp{N}\,J_{\alpha}\,\cos\,\theta_{\alpha}(t)\, \dot \varphi_{\alpha}(t).
\ee
Thus, our particle action is now given by
\be
S_{part}\sp{(N)}\,=\,S_{part,0}\sp{(N)}\,+\,\int \,dt\,\sum_{\alpha=1}\sp{N}\,
Q_{\alpha}\sp{a}(\dot x_{\alpha}A_{1,\alpha}\sp{a}\,+\,A_{0,\alpha}\sp{a})\,+\,S_{SU(2)}\sp{(N)}
\ee 
and the total action (see (1.2), (1.8) and (3.3)) by
\be
S\sp{(N)}\,=\,S_{field}\,+\,S\sp{NA}\,+\,S_{part}\sp{(N)}.
\ee

Looking at the equations of motion we note that for the zweibein fields we obtain, again, the Gauss constraint (2.4) and the non-Abelian analogue of (2.3) 
$(E\sp2\rightarrow (E\sp{a})\sp2$). We also have the corresponding requirements
at spatial infinity (the vanishing of $E\sp{a}$ and $F$ and finiteness of $e$ and $h$).
Therefore, in the gauge (2.7), the zweibein field $e(x,t)$ is again given by (2.8)
with the constraint (2.6). Moreover, analogously to (2.9) and (2.10) we obtain in the axial gauge $A_1\sp{a}=0$ that
\be
A_0\sp{a}\,=\,\frac{\kappa}{2} \,\sum_{\alpha}\,Q_{\alpha}\sp{a}\,\vert x-x_{\alpha}\vert
\ee
with the constraint
\be
\sum_{\alpha}\,Q_{\alpha}\sp{a}\,=\,0,
\ee
{\it ie} quantum mechanically we will have only isospin-singlet $N$-body states.
This corresponds to the fact that only gauge invariant, {\it i.e.} singlet, states are elements of the physical Hilbert-space (cp. [7]).

Along the same lines as in the Abelian case we conclude that the non-Abelian analogue to (2.3) is satisfied identically.

Finally, applying the Legendre transformation to the Lagrangian (3.4), using (2.8), (3.5)
and the constraints (2.6) and (3.6) we obtain for the two-body Hamiltonian 
$H$
\be
H\,=\,\frac{1}{4}\, {\dot x\sp2\over 1+\frac{\lambda}{2}\vert x\vert }\,+\,\frac{\kappa}{2}(Q\sp{a})\sp2\,\vert x\vert,
\ee
where we have defined
\be
Q\sp{a}:=\,Q_1\sp{a}\,=\,-Q_2\sp{a}.
\ee

Let us note that the Hamiltonian (3.7) has the same structure as in the Abelian case (2.12).
Therefore, we can give a common quantum mechanical treatment, for both cases,
and the conclusions will be the same (apart from the change of parameters).

\section{The Quantum Mechanical Two-Body Problem on a line}

When expressed in terms of canonical variables the classical Hamiltonian $H$
has the form
\be
H\,=\,p\sp2(1+\frac{\lambda}{2}\vert x\vert)\,+\,\frac{1}{2}q\vert x\vert
\ee
where $q$, a function of the particle charges, is given by
\be
q:=\cases{g\sp2\quad&\mbox{Abelian case}\cr \kappa(Q\sp{a})\sp2\quad &\mbox{non-Abelian case}\cr}.
\ee

In quantum mechanics the constraint (3.6) has to be considered as a subsidiary
condition on the wave function $\chi_2$
\be
(\hat Q_1\sp{a}\,+\,\hat Q_2\sp{a})\,\chi_2\,=\,0
\ee
{\it ie} the two particle states belong to the same isospin multiplet and couple to produce the total isospin zero.
Therefore, after quantization, we have in (4.2)
\be
(Q\sp{a})\sp2\rightarrow \hbar\sp2\tau(\tau+1)
\ee
with $\tau$ being the isospin of a single particle.

Solving the ordering problem involved in quantization of (4.1) by following the
prescription given in (\cite{five})\footnote{This prescription is in agreement with the Moyal quantization applied to the classical functions on phase space.} we obtain the following stationary
Schr\"odinger equation
\be
E\,\chi_2(x)\,=\,\left[-\hbar\sp2\,\partial_x\,(1+\frac{\lambda}{2}\vert x\vert)\partial_x\,-\,\frac{\hbar\sp2\lambda}{4}\delta(x)\,+\,\frac{1}{2}q\vert x\vert\right]\,\chi_2(x),
\ee
which differs from (1.5) by just the potential term $\frac{q}{2}\vert x\vert$ - due to the additional gauge interaction.

Now we proceed in complete analogy to (\cite{five}):
\begin{itemize}
\item
With Bose-symmetry $\chi_2(x)=\chi_2(-x)$ we obtain from (4.5) on $R_+\sp{1}$ the differential equation
\be
E\chi_2\,=\,\{-\hbar\sp2\partial_x\,(1+\frac{\lambda}{2}x)\partial_x\,+\,\frac{1}{2}qx\}\,\chi_2
\ee
with the boundary condition
\be
\partial_x\,\chi_2(0)\,=\,-\frac{\lambda}{8}\,\chi_2(0).
\ee
\item
We perform the change of variables
\be
x\,\rightarrow\,y:=\,\frac{4}{\vert\lambda\vert}\,\left(1+\frac{\lambda}{2}x\right)\sp{\frac{1}{2}}
\ee
and define
\be
\tilde \varphi_2(y):=\,(2+\lambda x)\sp{\frac{1}{4}}\,\chi_2(x)
\ee
to obtain the Schr\"odinger equation for $\tilde\varphi_2$

\begin{figure}[h]
\unitlength1cm
\hfil\begin{picture}(12,12)
%\put(2.5,5){What is this?}
\epsfxsize=11cm
\epsffile{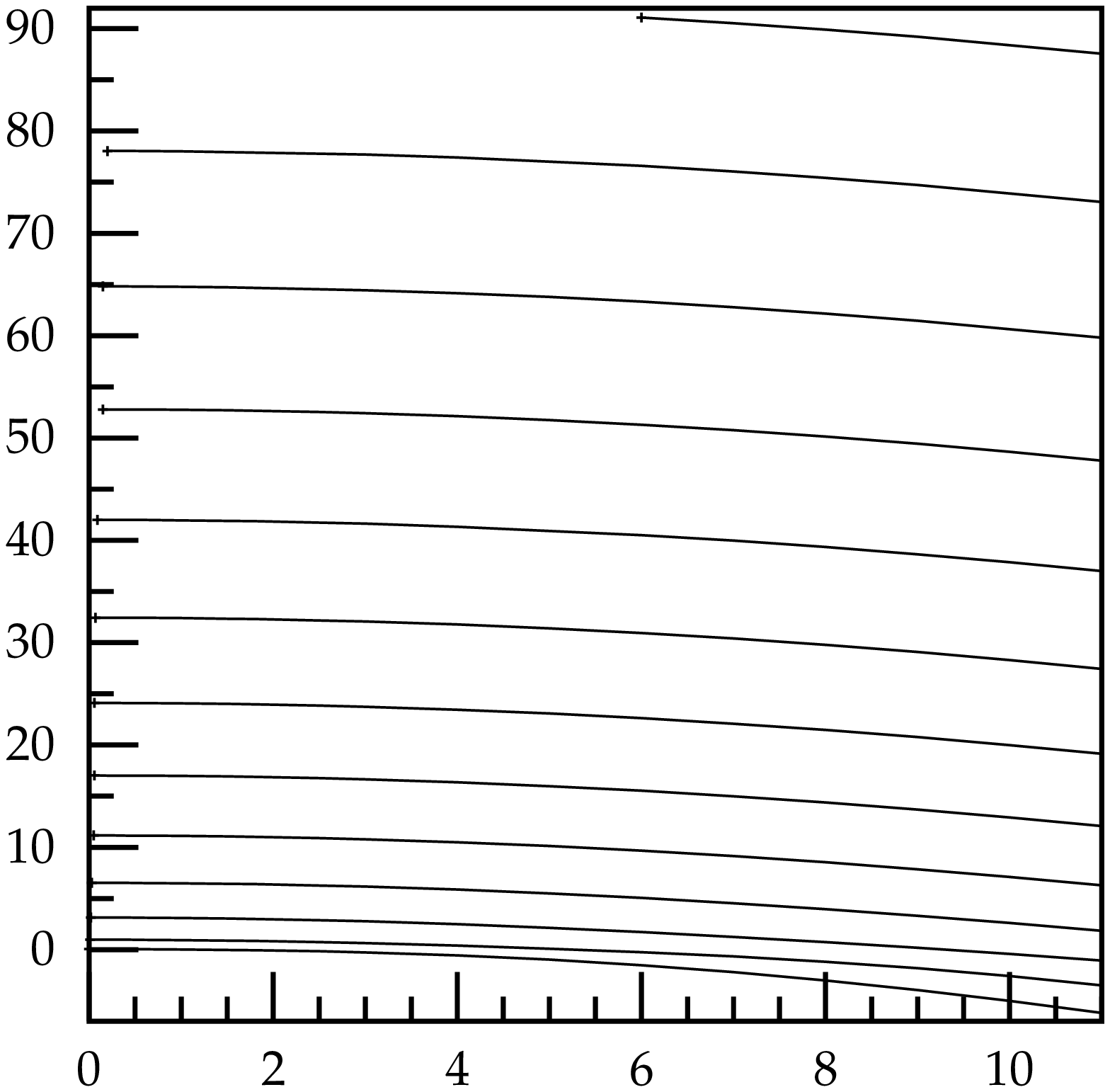}
\end{picture}
\caption{Energy (in units of
$\hbar\sp2\lambda\sp2$) as a function of $z={4\sqrt{\lambda q}\over \hbar \lambda\sp2}$}
\end{figure}

\be
E\tilde\varphi_2(y)\,=\,\left[-\hbar\sp2\partial_y\sp2\,-\,
\frac{\hbar\sp2}{4y\sp2}\,+\,\frac{q}{\lambda}\left(\frac{\lambda\sp2y\sp2}{16}
-1\right)\right]\,\tilde\varphi_2
\ee
together with the boundary condition
\be
\partial_y\,\tilde\varphi_2\left(\frac{4}{\vert\lambda\vert}\right)\,=\,0
\ee
\end{itemize}
In order to proceed further we have to consider separately the cases of $\lambda<0$ and $\lambda>0$.

\subsection{$\lambda<0$; Confinement by a Geometric Bag}

From (4.8) we see that $y\in (0,\,\frac{4}{\vert\lambda\vert})$. The requirement
of finiteness of $\chi_2$ at the edge of the bag ($x_0=\frac{2}{\vert\lambda\vert}$) leads by (4.9) to
\be
\tilde\varphi_2(0)\,=\,0.
\ee
The solution of (4.10) respecting the boundary condition (4.12), 
for a finite value of $\chi_2(\frac{2}{\vert\lambda\vert})$,
                    is given in terms of 
the confluent hypergeometric function $_1F_1$ by
\be
\tilde\varphi_2\,=\,z\sp{\frac{1}{4}}\,e\sp{-\frac{z}{2}}\,_1\!F_1(-A;1;z),
\ee
where we have defined
\be
z:=\,{\sqrt{\lambda q}\over 4\hbar }\,y\sp2
\ee
and
\be
A:=\,{\frac{q}{\lambda}\,+\,E \over \hbar \sqrt{\lambda q}}\,-\,{1\over 2}
\ee

\begin{figure}[h]
\unitlength1cm
\hfil\begin{picture}(12,12)
%\put(2.5,5){What is this?}
\epsfxsize=11cm
\epsffile{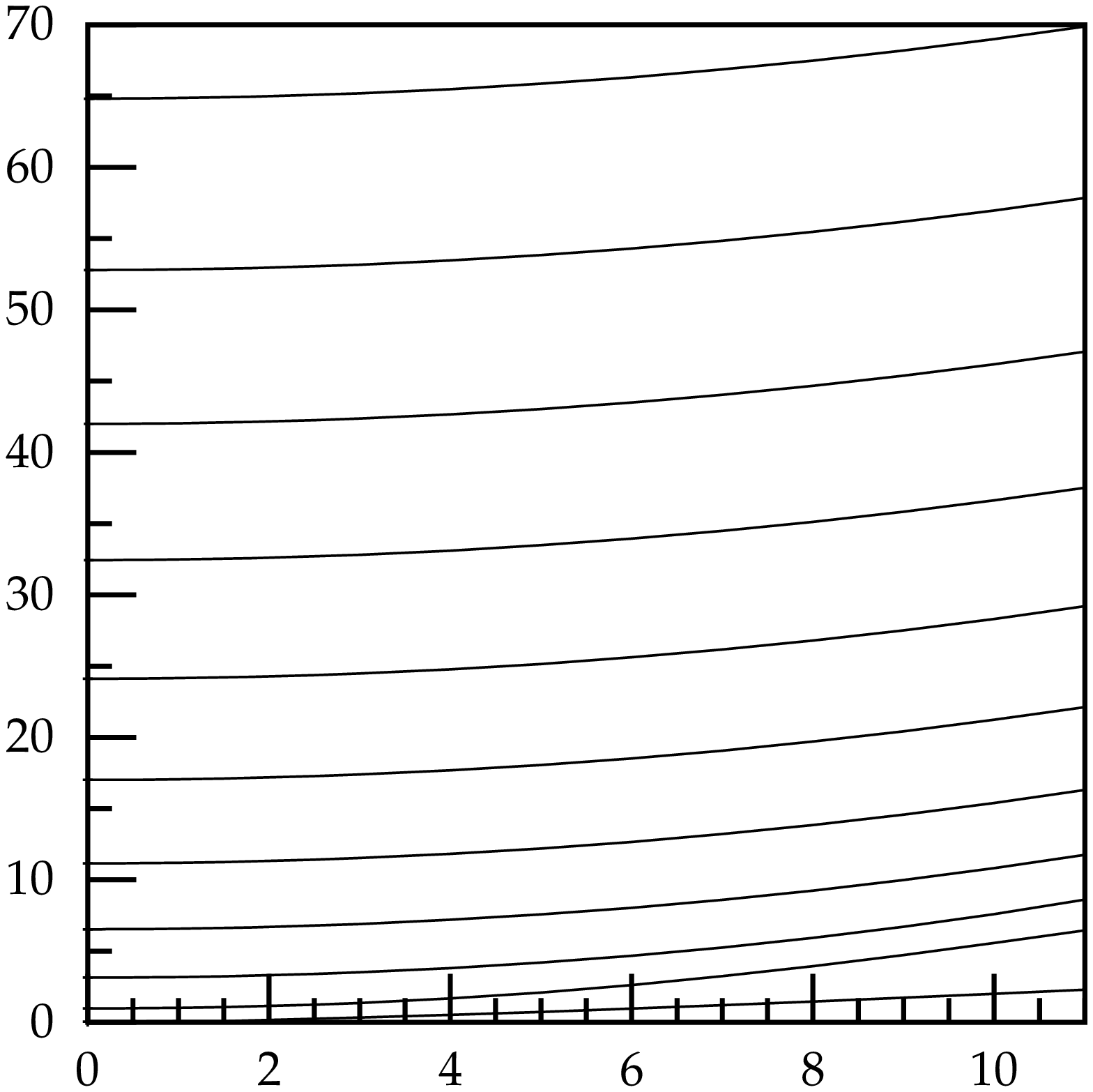}
\end{picture}
\caption{Energy (in units of
$\hbar\sp2\lambda\sp2$) as a function of $z={4\sqrt{\vert\lambda q\vert}\over \hbar\lambda\sp2}$}
\end{figure}

To determine the values of energies we have to resort to numerical methods.
We have performed such calculations for the lowest values of energies (as a function 
of $\lambda$ and $q$). To do this we have first observed that
(4.11) translates into the condition
\be
\,_1\!F_1\,(1-2z)\,+\,4z{\partial \,_1\!F_1\over \partial z}\,=\,0,
\ee
at $z={4\sqrt{\lambda q}\over \hbar \lambda\sp2}$, where $\,_1\!F_1\,=\,_1\!F_1(-A,1;z)$.
Thus defining $P$ as the left hand side of this formula
we have varied $A$ and determined the values of $A$ for which $P$ vanishes. This,
via (4.15) gives us the values of the energy.

Note that [8]
\be
{d\over dz}\,_1\!F_1(-A,1;z)\,=\,-A\,_1\!F_1(1-A,2;z)
\ee
and so $P$ is given by
\be
P\,=\,_1\!F_1(-A,1;z)(1-2z)\,-\,4zA\,_1\!F_1(1-A,2;z).
\ee

As $\lambda<0$ we note that $z$ is real for $q<0$.
 Of course, $q$ is given by (4.2) so it can 
be negative only in a nonabelian case.

We have performed a series of numerical calculations - determining 
zeros of $P$, as a function of $A$ for many values of $z$ and the results are
presented in Fig. 1.

\begin{figure}[h]
\unitlength1cm
\hfil\begin{picture}(12,12)
%\put(2.5,5){What is this?}
\epsfxsize=11cm
\epsffile{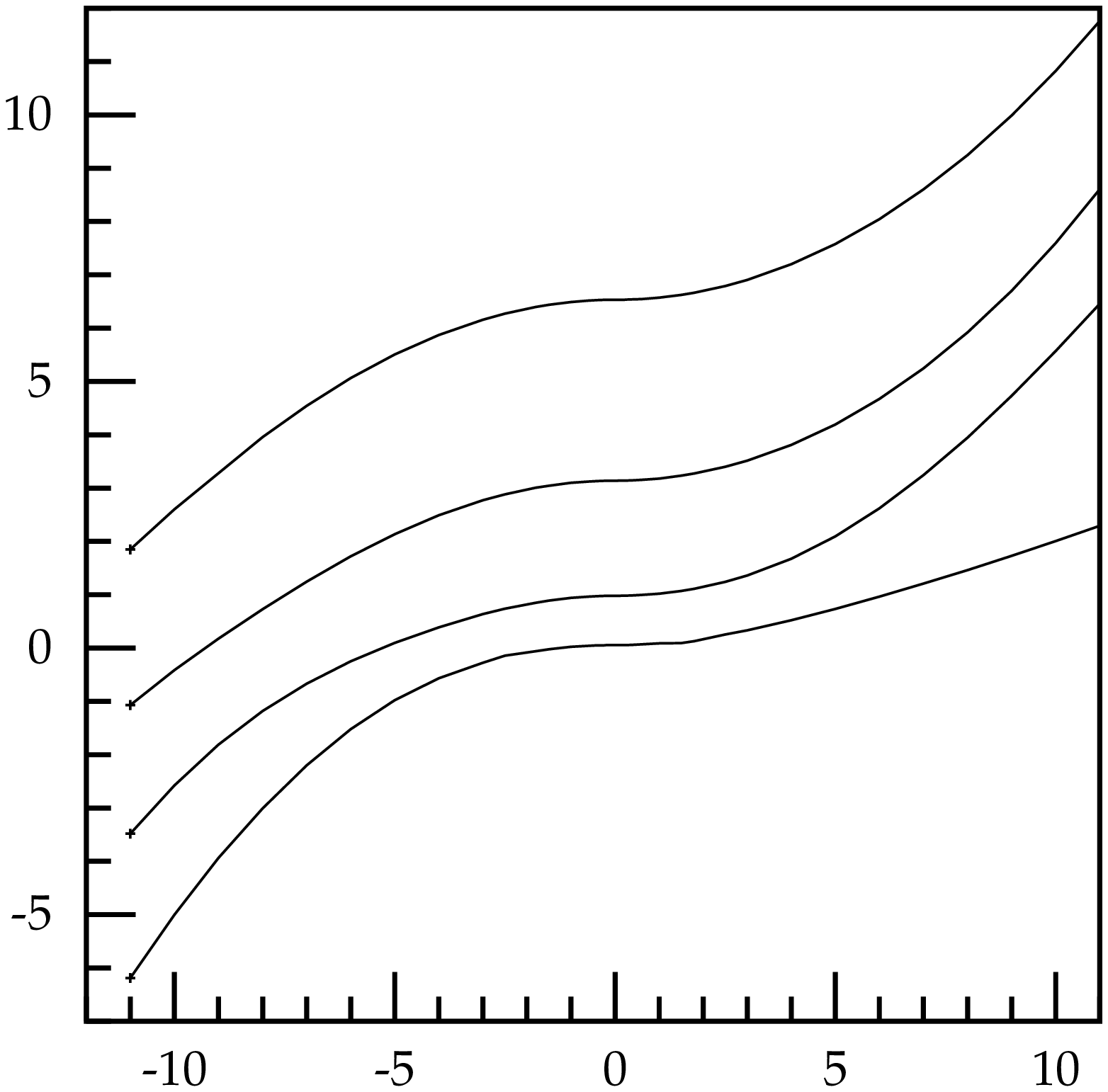}
\end{picture}
\caption{Energy (in units of
$\hbar\sp2\lambda\sp2$) as a function of $z={4\sqrt{\vert \lambda q\vert }\over \hbar \lambda\sp2}sign(q)$}
\end{figure}

In Fig. 1 the vertical axis gives $E$ (in units of $\hbar\sp2\lambda\sp2$), 
the horizontal $z={4\sqrt{\lambda q}\over \hbar \lambda\sp2}$ {\it ie} (4.14)
at $y={4\over \vert \lambda\vert}$.
Looking at the figure we see that (for each $z$) the energy is quantised and  
we note a relatively weak dependence on $z$ (with all
values of $E$ decreasing as $z$ increases).

Note that when $z\rightarrow 0$ and $A\rightarrow \infty$ ({\it ie} $q\rightarrow 0$) our equation
reduces to (cp. [5]) 
\be
2sJ_1(s)\,=\,J_0(s),
\ee
where 
$s={4\sqrt{E}\over \hbar \vert \lambda\vert}$
whose lowest solutions are
\be
E=0.05531,\quad 0.9798,\quad 3.1385,\quad 6,531,...
\ee
where $E$ is given in units of $\hbar\sp2\lambda\sp2$. 

For $q>0$ $z$ becomes complex and the calculations are more involved (as we need
to use complex functions {\it etc}). However, we have managed to determine
the dependence of the few lowest energies on $q$. 
Interestingly and reassuringly, both the real and imaginary
parts of the complex function $\,_1\!F_1$ vanished at the same value of $A$.
 Our results
are presented in Fig. 2, where on the horizontal axis we have put
$z={4\sqrt{\vert \lambda q\vert }\over \hbar\lambda\sp2}$.

We note that, as before, there is little dependence on $z$ but, this time,
the values of the energy increase with an increase of $z$.

In Fig. 3 we put together 4 lowest energies as functions of $z$ for 
both positive and negative values of $q$; hence the horizontal axis
now shows $z={4\sqrt{\vert \lambda q\vert }\over \hbar \lambda\sp2}sign(q)$.

It is very interesting to note that for a particular value of $z$, namely
$z={1\over 2}$ which corresponds to $q={\hbar\sp2\lambda\sp3\over 64}$,
our model fits into the scheme of supersymmetric quantum mechanics\footnote{see
{\it e.g} the review article [9] and the literature cited therein.}.
At $z=\frac{1}{2}$ the Hamiltonian (4.10) factorizes as
\be
H\,=\,B\sp{\dagger}B\,+\,E_0\,=\,H_1\,+\,E_0
\ee
with
\be
B\,:=\,\hbar \partial_y\,+\,W(y),\qquad B\sp{\dagger}\,:=\,-\hbar \partial_y\,+\,W(y)
\ee
and the superpotential
\be
W(y)\,:=\,{\hbar \lambda\sp2\over 32}y\,-\,{\hbar\over 2y},
\ee
where
\be
E_0\,=\,\frac{3}{64}\hbar\sp2\lambda\sp2
\ee
is the ground state energy of $H$.

The supersymmetric partner of $H_1$ is given by
\be
H_2\,:=\,BB\sp{\dagger}\,=\,-\hbar\sp2\partial\sp{2}_y\,+\,\left({\lambda\sp2\hbar
\over 32}\right)\sp2y\sp2\,+\,\frac{3}{4}{\hbar\sp2\over y\sp2}.
\ee
Note that the ground-state wave function of $H_1$
\be
\Psi_0\sp{(1)}\,=\,\exp\left(-\frac{1}{\hbar}\,\int \,dy\,W(y)\right)
\ee
satisfies the boundary condition (4.11) due to
\be
W(\frac{4}{\vert\lambda\vert})\,=\,0
\ee
so the supersymmetry is unbroken.
Thus $H_2$ has the same spectrum as $H_1$ with the exception
of the ground state [9]:
\be
E_{n+1}\sp{(1)}\,=\,E_n\sp{(2)},\quad n\in N,
\ee
with 
\be
\Psi_n\sp{(2)}\,\sim\,B\,\Psi_{n+1}\sp{(1)}.
\ee

Using (4.11) and (4.27) we obtain from (4.29) the boundary
condition
\be
\Psi_n\sp{(2)}\left(\frac{4}{\vert\lambda\vert}\right)\,=\,0.
\ee
Therefore, the excitation energies of $H$ can be read off from (4.30)
which, due to (4.25), takes the form:
\be
\quad_1\!F_1\left(1-A,2;\frac{1}{2}\right)\,=\,0,
\ee
which agrees with (4.18) taken at $z=\frac{1}{2}$ for $A\ne 0$.

Note that the roots of (4.31) are approximately given [8] by
\be
A_n\,=\,\frac{\pi\sp2}{2}\left(n+\frac{1}{4}\right)\sp2,\quad n=1,2.3....
\ee 
in good agreement with our, numerically determined, values.

Finally, let us note that our bag is impenetrable both classically and
quantum mechanically. The boundary condition (4.12) allows no
Schr\"odinger current at \qquad
$x_0=\frac{2}{\vert\lambda\vert}$ and therefore no tunneling through 
the edge of the bag.

\subsection{$\lambda>0$; Confining oscillator potential}

For $\lambda>0$ $y$ lies on the half-axis (${4\over \lambda},\,\infty$).

\begin{figure}[h]
\unitlength1cm
\hfil\begin{picture}(12,12)
%\put(2.5,5){What is this?}
\epsfxsize=11cm
\epsffile{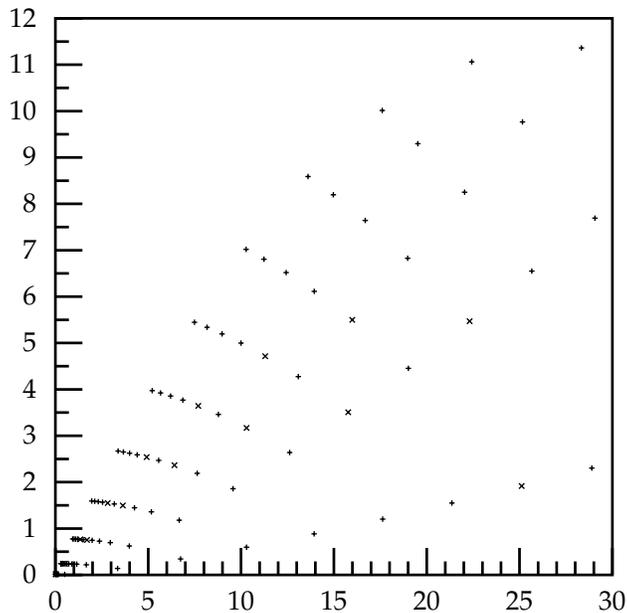}
\end{picture}
\caption{Some values of energy (in units of
$\hbar\sp2\lambda\sp2$) as a function of $z={4\sqrt{\lambda q}\over \hbar \lambda\sp2}$}
\end{figure}

We look for solutions of the Schr\"odinger equation (4.10) for $q>0$ with $\tilde\varphi_2\in L\sp2({4\over \lambda},\infty)$ obeying the boundary condition (4.11). This defines a self-adjoint eigen-value problem leading to a discrete energy spectrum which is bounded from below. 
To be more specific, we note that we have to replace $\,_1\!F_1$ of section 
4.1 by Kummer's function $U(-A,1;z)$, which at infinity grows only as a power
($\sim z\sp{A}$) and so, due to (4.13) guarantees the square integrability.
In this case the numerical calculations of zeros of $P$ are somewhat 
cumbersome and so we have not carried them out.

Nevertheless we can obtain some energy values for a discrete set of $z$ values
by the following argument:

We start with the defining relation between the Kummer $U$ function and
the $\,_1\!F_1$ functions [8]:
\be
U(-A,1;z)\,=\,\lim_{b\rightarrow 1}{\pi\over \sin\,\pi b}
\left\{{\,_1\!F_1(-A,b;z)\over \Gamma(1-b-A)\Gamma(b)}
-z\sp{1-b}{\,_1\!F_1(1-b-A,2-b;z)\over \Gamma(-A)\Gamma(2-b)}\right\}.
\ee

Thus, for $A$ equal to a positive integer $n$ we obtain
\be
U(-n,1;z)=\lim_{b\rightarrow 1}{\Gamma(1-b)\over \Gamma(1-b-n)}\,_1\!F_1(-n,b;z)
= (-1)\sp{n}\,n!\,_1\!F_1(-n,1;z).
\ee
But we have [8]
\be 
\,_1\!F_1(-n,1;z)\,=\,L_n(z),
\ee
where $L_n(z)$ are the Laguerre polynomials.

\begin{figure}[h]
\unitlength1cm
\hfil\begin{picture}(10,10)
%\put(2.5,5){What is this?}
\epsfxsize=8cm
\epsffile{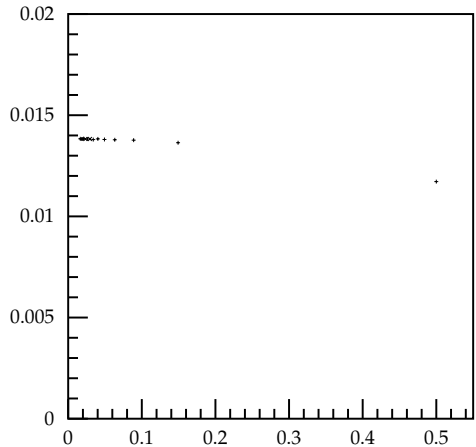}
\end{picture}
\caption{Lowest values of energy (in units of
$\hbar\sp2\lambda\sp2$) as a function of $z={4\sqrt{\lambda q}\over \hbar \lambda\sp2}$}
\end{figure}

Thus we see that for those values of $z$ which are roots of $P$ (4.18) and
fulfill the relation
\be
A(z)\,=\,n
\ee
our problem reduces mathematically to the well known case of a charged particle moving in a $s$-state on a plane and interacting with a constant
magnetic field $B$ ({\it ie} - the Landau problem). 
The particular $z$ values satisfying (4.36) can be read off from Fig.1

In Fig. 4 we plot some values of the energies determined this way.
We plot only those corresponding to $z={4\sqrt{ \lambda q }\over \hbar \lambda\sp2}<30$ and determined from $n\le 13$. Assuming that the extrapolation to non-integer values of $A$ would give similar results
(and there is no reason to expect it to be otherwise) we see that, at every value of $z$ there is a tower of energies. Moreover, these energies
decrease as $z$ increases.
The lowest energy, which is not really visible in Fig. 4) is plotted
in Fig. 5. In this case, our approach gave us results in a very small range
of $z$, and all the results show very little dependence on $z$.

\section{Conclusions}
In the present paper we have considered the confinement mechanisms for two
nonrelativistic particles on a line arising from the addition to the non-standard
gravity [5] of an additional (non)-Abelian gauge interaction.
Our results show that for
\begin{itemize}
\item $\lambda<0$ and for any sign of the additional gauge coupling $q$ we
observe confinement by the geometric bag formation mechanism with
 only weak dependence of the energy spectrum on $q$.
\item $\lambda>0$ the confinement is due to the rising potential term.
\end{itemize}

Note that the confinement found for $\lambda>0$ is the well known confinement
mechanism of two-dimensional gauge theories ([3],[4]). Addition of nonstandard 
gravity only alters the energy spectrum as a function of the coupling constant
$\lambda$. 

The confinement found for $\lambda<0$ is, however, of a completely
different nature. It arises, selfconsistently, from a singularity of the nontrivial metric determined dynamically by the nonstandard gravity interaction
of the confined particles. This is in contrast with the current treatments
of (3+1)-dimensional Yang-Mills theories for which it is only the gauge fields 
that form a geometric bag which confines test particles (cp. [10] and the review [11]). The results of the present paper,  together with the corresponding
results in (2+1)-dimensions with ([12]) or without ([6])  an additional
(non)-Abelian gauge interaction, strengthen our feeling that nonstandard
gravity might be of some relevance for the solution of the confinement
problem in strong interactions. Thus, further research into nonstandard gravity,
in particular in (3+1)-dimensions, is called for.

\textheight 8.8in \textwidth 6in

\end{document}